\documentclass[conference,10pt]{IEEEtran}
\IEEEoverridecommandlockouts

\usepackage[utf8]{inputenc}
\usepackage{microtype}
\usepackage[colorlinks=true,urlcolor=blue,citecolor=blue,linkcolor=blue]{hyperref}
\usepackage{cite}
\usepackage{braket}
\usepackage{algorithm}
\usepackage{verbatim}
\usepackage{mdframed}
\usepackage{algpseudocode}
\usepackage{amsmath,amssymb,amsfonts}
\usepackage{graphicx, xcolor}
\usepackage{textcomp}

\begin{document}

\title{
Optimizing Ansatz Design in Quantum Generative Adversarial Networks Using Large Language Models
}

\author{\IEEEauthorblockN{Kento Ueda}
\IEEEauthorblockA{\textit{IBM Quantum} \\
\textit{IBM Research -- Tokyo} \\
Tokyo, Japan \\
Kento.Ueda@ibm.com}
\and
\IEEEauthorblockN{Atsushi Matsuo}
\IEEEauthorblockA{\textit{IBM Quantum} \\
\textit{IBM Research -- Tokyo} \\
Tokyo, Japan \\
matsuoa@jp.ibm.com}
}

\maketitle

\begin{abstract}
We present a novel approach for improving the design of ansatzes in Quantum Generative Adversarial Networks (qGANs) by leveraging Large Language Models (LLMs). By combining the strengths of LLMs with qGANs, our approach iteratively refines ansatz structures to improve accuracy while reducing circuit depth and the number of parameters. 
This study paves the way for further exploration in AI-driven quantum algorithm design. The flexibility of our proposed workflow extends to other quantum variational algorithms, providing a general framework for optimizing quantum circuits in a variety of quantum computing tasks.

\end{abstract}

\begin{IEEEkeywords}
Quantum Machine Learning, qGAN, Ansatzes, LLMs
\end{IEEEkeywords}

\section{Introduction}
Quantum computing has been gathering significant attention for its potential to solve complex tasks such as integer factorization~\cite{shor}. However, current quantum devices face challenges such as high error rates and limitations in the size of executable quantum circuits~\cite{preskillNISQ}, which hinders the execution of quantum circuits for more complicated tasks. 
Despite these challenges, researchers are actively exploring ways to optimize the use of existing quantum computers~\cite{kim2023evidence}.

One particularly promising area of research is quantum machine learning, which aims to harness the capabilities of current quantum devices~\cite{gujju2024quantum}.
Among the various approaches in this field, Quantum Generative Adversarial Networks (qGANs) stand out. qGANs extend the classical concept of Generative Adversarial Networks (GANs) into the quantum domain~\cite{lloyd2018quantum, electronics12040856}. 
By leveraging quantum mechanics, qGANs enhance the generative capabilities and efficiency of GANs. In qGANs, the generator and/or discriminator are implemented using quantum circuits, potentially offering exponential speed-ups in training and data generation compared to classical GANs.
Applications of qGANs are diverse, ranging from generating quantum states~\cite{dallaire2018quantum, benedetti2019adversarial, hu2019quantum} to producing classical distributions~\cite{situ2020quantum, romero2021variational, zeng2019learning, zoufal2019quantum}. 
Moreover, qGANs can be combined with other quantum algorithms. Using the data prepared by qGANs in algorithms such as Quantum Amplitude Estimation (QAE)~\cite{brassard2002quantum} or the HHL-algorithm~\cite{harrow2009quantum} can help achieving quantum advantage.

While the potential of qGANs is vast, there remains room for improvement, particularly in the design of the ansatz within qGANs.
This challenge is similar to those encountered in other variational quantum algorithms like the Variational Quantum Eigensolver (VQE)~\cite{peruzzo2014variational}.
Currently, the ansatzes employed in qGANs typically have static structures, which may limit their performance. There are a few approaches that dynamically change the structures of ansatzes.
\cite{grimsley2019adaptive, rattew2019domain, matsuo2023enhancing}

Machine learning has proven to be a valuable tool in this regard, as it offers techniques that can enhance the efficiency and performance of quantum computation~\cite{kremer2024practical, jaouni2024deep, liao2023machine,krenn2023artificial}.
In machine learning, Large Language Models (LLMs) have shown extraordinary success across a variety of domains ~\cite{van2023chatgpt, ouyang2022training}, far beyond their initial application in natural language processing. Their ability to explore complex spaces that are challenging for humans to navigate makes them a powerful tool for optimizing quantum systems.

In this paper, we propose a novel workflow that leverages LLMs to iteratively refine ansatz designs, enhancing accuracy while simultaneously reducing both circuit depth and the number of parameters.
There are similar studies that uses LLMs to design ansatzes conducted independently of ours~\cite{liang2023unleashing, nakaji2024generative}. 
In~\cite{liang2023unleashing}, they construct the ansatz by combining small two-qubit gate blocks and uses only the final value of a single metric as feedback information. In contrast, our proposed method constructs the ansatz by combining larger sub-circuit layers and incorporates various data, including intermediate values, as input during feedback.
In~\cite{nakaji2024generative}, they consider only gates without parameters.
This paper focuses on qGANs, but our proposed framework is also applicable to more general variational algorithms.

The remainder of this paper is structured as follows: Section~\ref{sec:background} provides background information on quantum circuits
and qGANs. Section~\ref{proposed} presents our proposed approach. 
Finally, Section~\ref{sec:discussion} concludes the paper and highlights
potential avenues for future research.

\section{Background}
\label{sec:background}
In this section, we provide an overview of quantum circuits and qGANs.

\subsection{Quantum Circuits}
Quantum circuits are fundamental models for quantum computation. They consist of qubits and a series of quantum gates~\cite{nielsen}.
In contrast to classical bits, which are limited to states 0 or 1, qubits in quantum computing can be in the state $\ket{0}$, $\ket{1}$, or a superposition of both. This superposition state is a linear combination of $\ket{0}$ and $\ket{1}$, expressed as $\alpha \ket{0} + \beta \ket{1}$, where $\alpha$ and $\beta$ are complex numbers and satisfy the condition $|\alpha|^2 + |\beta|^2 = 1$. These coefficients, $\alpha$ and $\beta$, are referred to as the amplitudes of the respective basis states. An $n$-qubit state can be described by $\ket{\psi} = \sum_{k \in {0,1}^n} \alpha_k \ket{k}$, where each $\alpha_k$ is a complex number and the sum of their magnitudes squared equals 1. This state can alternatively be represented as a $2^n$-dimensional vector, such as $(\alpha_0, \alpha_1, \ldots, \alpha_{2^n -1})^T$.
Quantum gates are responsible for executing specific unitary operations on qubits. They determine which unitary operation is applied to the corresponding qubits, and the operations are performed sequentially from left to right.

\subsection{Quantum Generative Adversarial Networks}
GANs are a class of machine learning frameworks~\cite{goodfellow2014generative}. GANs consist of two neural networks, the generator and the discriminator, which are trained simultaneously through adversarial processes. 
During training, the generator improves its ability to produce realistic data, and the discriminator becomes more skilled at detecting fakes. 

The objective function of GANs is formulated as a minimax game between the generator $G$ and the discriminator $D$. 
The objective function is expressed as:
\begin{equation}
\min_G \max_D \mathbb{E}_{\mathbf{x} \sim p_{\text{data}}(\mathbf{x})} [\log D(\mathbf{x})] + \mathbb{E}_{\mathbf{z} \sim p_{\mathbf{z}}(\mathbf{z})} [\log (1 - D(G(\mathbf{z})))]
\label{eq:gan_objective}
\end{equation}
where \( \mathbb{E} \) denotes the expected value, \( \mathbf{x} \) represents samples from the real data, \( p_{\text{data}}(\mathbf{x}) \) is the distribution of the real data, \( \mathbf{z} \) represents samples of random noise input to the generator, \( p_{\mathbf{z}}(\mathbf{z}) \) is the distribution of the random noise, \( D(\mathbf{x}) \) is the probability that the discriminator correctly identifies real data \( \mathbf{x} \) as real, \( G(\mathbf{z}) \) is the data generated by the generator from the noise \( \mathbf{z} \), and \( D(G(\mathbf{z})) \) is the probability that the discriminator identifies the generated data as real.
The discriminator aims to maximize the probability of correctly identifying real data and minimize the probability of mistakenly identifying generated data as real. Conversely, the generator aims to minimize the probability that the discriminator correctly identifies the generated data as fake.

qGANs extend the concept of classical GANs into the realm of quantum computing~\cite{lloyd2018quantum}. 
qGANs leverage the principles of quantum mechanics to enhance the generative capabilities and efficiency of GANs. In qGANs, the generator and/or discriminator can be implemented using quantum circuits.

In this paper, we utilize qGANs configuration featuring a quantum generator paired with a classical discriminator.
The quantum generator is implemented as a variational quantum circuit, known as an ansatz, with tunable parameters, which are optimized during the training process. Detailed implementation specifics are provided in Section~\ref{proposed}.

\section{Large Language Models-Guided Quantum Ansatz Search (LLM-QAS)}
\label{proposed}

The LLM understands the context through interaction with the user and provides responses based on that understanding. It can handle complex instructions and nuances from the user, and it improves the accuracy of its responses by incorporating feedback during the conversation. LLM-based models have demonstrated remarkable utility across various research fields ~\cite{van2023chatgpt, ouyang2022training}. We propose a workflow to search for optimal ansatzes suited for specific problems using LLMs.

\begin{figure*}[tb]
\centering
\includegraphics[width=1.0\columnwidth]{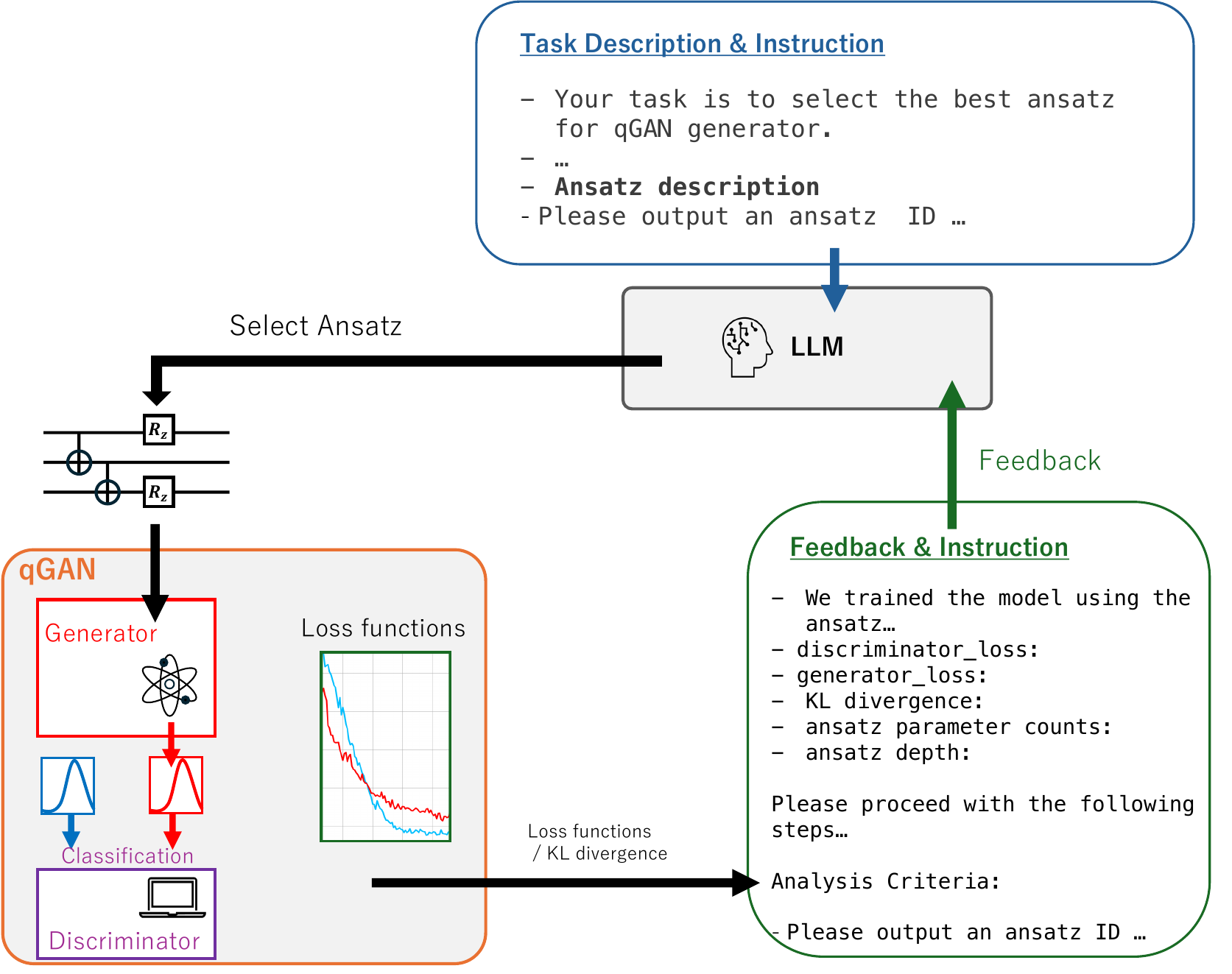}
\caption{Workflow using Large Language Models-Guided Quantum Ansatz Search (LLM-QAS). An ansatz generated by the LLM, based on a prompt containing task descriptions and instructions, is provided as input to the generator of the qGAN. The loss function values and hyperparameters of the ansatz are then incorporated into a feedback prompt. The LLM generates a new ansatz by considering this feedback.}
\label{fig:workflow}
\end{figure*}

Fig.~\ref{fig:workflow} illustrates the LLM-QAS workflow, which optimizes the ansatz for the generator of a qGAN by leveraging an LLM. The workflow proceeds as follows:

\textbf{1) Task Description and Instructions:} The LLM is assigned to the task of selecting the optimal ansatz for the qGAN. Specific instructions provide the necessary information for the LLM to search for the best ansatz. This includes explanations or code for the ansatz candidates, as well as details on ansatz depth, parameter counts, and relevant statistical indicators. The LLM is instructed to execute the task and output an ansatz to be used for qGAN training.

\textbf{2) Ansatz Selection:} Based on the provided instructions, the LLM selects combinations of ansatzes that consist of the ansatz candidates (see Fig.~\ref{fig:ansatzes}). This selected ansatz combination is then applied to the qGAN generator.

\textbf{3)	qGAN Training and Evaluation:} The qGAN is trained using the selected ansatz combination on a quantum computer or a quantum simulator. During training, key metrics such as the discriminator and generator loss values, along with the Kullback-Leibler (KL) divergence~\cite{kullback1951information}, which quantifies the difference between two probability distributions, are tracked.

\textbf{4)	Feedback and Further Instructions:} Once the qGAN training is completed, feedback on performance metrics (discriminator loss, generator loss, KL divergence, ansatz parameter counts, and ansatz depth) is provided to the LLM. The LLM analyzes the feedback and refines the ansatz selection process accordingly.

\textbf{5)	Iterative Ansatz Refinement:} Using the feedback, the LLM iteratively selects and adjusts the ansatz to achieve better performance metrics, repeating the process as needed to optimize qGAN performance.

This LLM-QAS workflow demonstrates that it is effective to guide the selection and optimization of quantum ansatz architectures through an iterative feedback loop, leveraging the capability of the LLM interaction. By continuously refining the ansatz based on empirical performance data, this workflow significantly improves the qGAN training process, resulting in superior model performance.

We focus on roles of each component in the following subsections.

\subsection{Task Description prompt}
We first need to provide the LLM with a task description and the necessary requirements in the form of a prompt. The goal of this paper is to use a qGAN to load a quantum state that represents a target probability distribution. The LLM will search for the quantum ansatz that generates this quantum state. The prompt should include the following elements: an overview of the task, relevant code, an explanation of the qGAN training procedure, a description of the ansatz candidates (as shown in Fig.~\ref{fig:ansatzes}), and, finally, instructions for the desired output from the LLM. The prompt is shown in Appendix ~\ref{prompt1}.

\subsection{Ansatz Candidates}
In the LLM-QAS workflow, the LLM is provided with several ansatz candidate options and is tasked with proposing combinations of these as the ansatz used for the qGAN generator. Notably, the LLM is permitted to choose duplicate ansatz candidates. Alternative approaches to leveraging an LLM to search for suitable ansatzes could include designing an ansatz from scratch or providing a set of gate candidates, within which the LLM must design the ansatz.

\begin{figure}[tb]
\centering
\includegraphics[width=0.9\columnwidth]{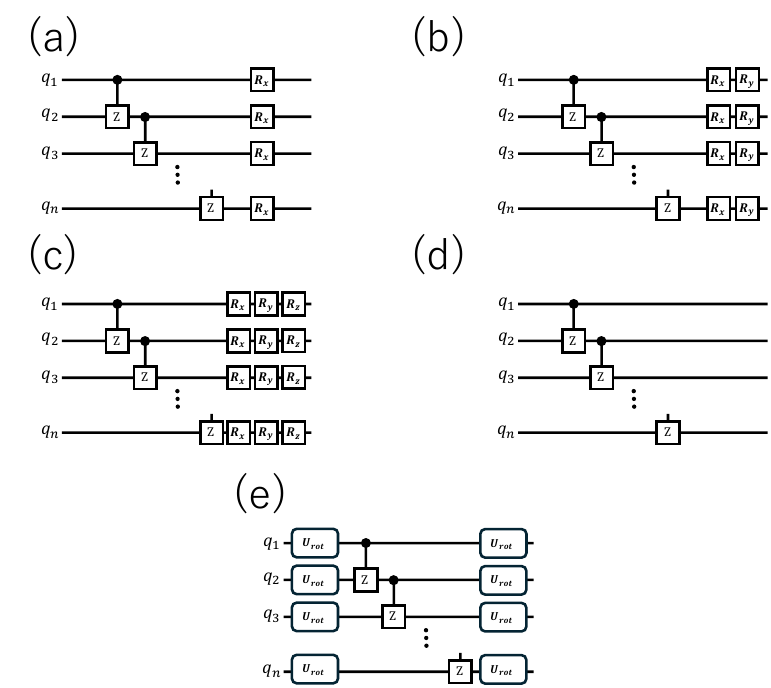}
\caption{Ansatz Candidates used in the LLM-QAS workflow. From (a) to (e), the ansatz candidates used in the LLM-QAS are shown.}
\label{fig:ansatzes}
\end{figure}

The five ansatz candidates employed are shown in Fig.~\ref{fig:ansatzes}: (a) $CZ$ and $RX$, (b) $CZ$ and ($RX$, $RY$), (c) $CZ$ and ($RX$, $RY$, $RZ$), (d) $CZ$ only, and (e) $TwoLocal$. These ansatz blocks are commonly used in problems involving parameterized ansatzes. By providing the LLM with user-defined ansatz candidates, we need to restrict the search space, increasing the likelihood of finding an optimal solution with fewer LLM-QAS workflow iterations. The $TwoLocal$ ansatz, as described in the documentation of \cite{qiskit2024}, differs from other ansatzes in that it requires additional configuration options when selected. When the $TwoLocal$ ansatz is chosen, the LLM is instructed to specify further detailed ansatz settings. Specifically, the LLM must choose an entanglement strategy from $Full$, $Linear$, $Reverse\_linear$, $Pairwise$, $Circular$, and $SCA$, which define qubit interactions. As the entanglement strategy, $Full$ entanglement connects every qubit with all others, maximizing connectivity. $Linear$ entanglement connects each qubit with its neighbor in sequence, while $Reverse\_linear$ entanglement does the same in reverse order. $Pairwise$ entanglement alternates connections between even and odd qubits across two layers. $Circular$ entanglement extends the $Linear$ entanglement model by linking the first and last qubits. $SCA$ (Shifted-Circular-Alternating) entanglement is a variation where the first-last qubit connection shifts with each block, and control-target roles alternate~\cite{sim2019expressibility}. The LLM also determines the single-qubit rotation gates to be used from $RX$, $RY$, and $RZ$ gates. The selection of the $TwoLocal$ ansatz allows the LLM to represent more complex ansatz structures.

\subsection{qGAN training}
The architecture of the qGAN consists of a hybrid quantum and classical system. We can implement the qGAN architecture using Qiskit libraries such as Qiskit-Aer, Qiskit Machine Learning, and Qiskit Algorithm~\cite{qiskit2024}. 
We can train the qGAN for several target distributions in the process of the LLM-QAS workflow.

We need to prepare the ansatz for the generator of the qGAN. First, we apply the Hadamard gates to the initial qubit state, followed by the LLM-QAS generated ansatz. The discriminator is implemented as a classical neural network using PyTorch~\cite{paszke2019pytorch}, with an input layer of size equal to the input data dimension. The architecture consists of five hidden layers with $256$, $128$, $64$, $32$, and $16$ nodes, respectively. Each hidden layer applies a linear transformation followed by a Leaky ReLU activation function with a negative slope of $0.2$, Batch Normalization, and a dropout of $0.3$. The final output layer applies another linear transformation to produce a single node output, followed by a sigmoid activation function. The topology of the discriminator was chosen based on empirical testing. The qGAN is trained using AMSGRAD~\cite{reddi2019convergence} with an initial learning rate set to $10^{-4}$. The training data is shuffled and divided into batches of $2000$, but only a single batch is used in real-device experiments due to time constraints. As the log-likelihood loss function does not necessarily reflect convergence, the KL divergence is employed as an appropriate evaluation metric. Each training setting is repeated $10$ times to assess the robustness of the results.

\subsection{Feedback prompt}
In the LLM-QAS workflow, the LLM proposes the next ansatz based on the training results of the qGAN and the current ansatz used for the generator. To guide the LLM in suggesting improved ansatz structures, the following information is provided: the loss function values from the qGAN training, the KL divergence as a statistical measure, the number of parameterized gates in the ansatz, and the ansatz depth. This information is incorporated into the feedback prompt (see Fig.~\ref{fig:workflow}), along with further instructions for the LLM. The feedback prompt is shown in Appendix ~\ref{prompt1}.

\section{Discussion}
\label{sec:discussion}
In this paper, we introduced LLM-QAS, a novel workflow for optimizing ansatz selection in quantum Generative Adversarial Networks (qGANs). By integrating large language models (LLMs) with quantum computational frameworks, LLM-QAS demonstrates the potential of leveraging LLMs to guide the iterative refinement of quantum ansatzes. The workflow's feedback loop enables systematic optimization of key parameters, such as ansatz depth and gate count, while preserving high model performance, even at larger qubit scales.

Ansatzes optimized using the LLM-QAS workflow in a quantum simulation environment can be directly executed on a quantum computer. This approach suggests that optimizing ansatzes on simulators and transferring them to quantum hardware is both practical and effective.

This study paves the way for further exploration of AI-driven quantum algorithm design, such as quantum finance and quantum machine learning. Moreover, the flexibility of the LLM-QAS workflow can be applied to other quantum variational algorithms, offering a generalizable framework for optimizing quantum circuits across diverse computational tasks.

Future work will focus on deploying LLM-QAS on quantum hardware to fully realize its potential. Expanding the ansatz candidate space and refining feedback prompts will also be critical for tackling more complex quantum problems and advancing the capabilities of AI-driven quantum algorithm design.

\small

\bibliographystyle{IEEEtran}

\newpage
\onecolumn
\appendix
\renewcommand{\thefigure}{A.\arabic{figure}}
\setcounter{figure}{0}

\subsection{Prompts used in LLM-QAS}
We provide full prompts as following.
\begin{center}
\label{prompt1}
Prompt: Task Description and Instruction
\begin{mdframed}
\begin{verbatim}
Your task is to help select the best ansatz for quantum Generative Adversarial Networks 
(qGANs) generator. 
The qGAN is a hybrid quantum-classical algorithm used for generative modeling tasks. 
It combines a quantum generator, which is an ansatz (parametrized quantum circuit), 
and a classical discriminator, a neural network, to learn the underlying probability 
distribution from the training data.

The generator and discriminator are trained in alternating optimization steps. 
The generator aims to produce probabilities that the discriminator will classify as 
training data values (i.e., probabilities from the real training distribution). 
The discriminator, on the other hand, tries to differentiate between the original 
distribution and the probabilities generated by the generator, 
effectively distinguishing between real and generated distributions. 
The goal is for the quantum generator to learn a representation of 
the target probability distribution.

Once trained, the quantum generator can be used to load a quantum state 
that approximates the target distribution.

==Classical discriminator==

We define a PyTorch-based classical neural network that represents the classical 
discriminator. The underlying gradients can be automatically computed with PyTorch.

{Discriminator code HERE}

==Loss function and entropy==

We will train the generator and the discriminator with binary cross entropy as the loss 
function:

{Definition of the loss function and its code HERE}


The final trained data and real data are evaluated by scipy.stats.entropy.

==Model Training workflow==

The model training flow is as follows:
In the training loop, we monitor not only the loss functions but also the relative entropy. 
The relative entropy describes a distance metric for distributions, allowing us to benchmark 
how close or far the trained distribution is from the target distribution.

{Model Training code HERE}

==Quantum ansatz for the generator==

- Number of qubits: <Number of qubits>
- Default number of circuit blocks: <Circuit block>

For each block, there are <Number of ansatz candidates> types of ansatz 
candidates to choose from:

{Description of the ansatz candidates and code HERE}


==INSTRUCTION ==

Please output an ansatz number as well as the selected qubits for each block. 
For example, [1,2,3,4] means we use operation 1 for block 1 and the block 1 is 
on all the qubits, operation 2 for block 2 and block 2 is on all the qubits, 
and so on, until operation 4 for block 4 and block 4 is on all the qubits.
Also, you can choose the same ansatz even if you have already selected the 
ansatz for the previous block.

Please output the values like this:

```
improved_ansatz_list = [4,1,5,1]
```

\end{verbatim}
\end{mdframed}
\end{center}

\begin{center}

Prompt: Feedback
\begin{mdframed}
\begin{verbatim}
We trained the model using the ansatz you provided. Below are the tracked values:

discriminator_loss_values: {discriminator loss values}
generator_loss_values: {generator loss values}
entropy_values: {entropy values}
ansatz parameter: {circuit parameter counts}
ansatz depth : {circuit depth}

Please proceed with the following steps:

1.	Train the quantum ansatz using the provided cost function code.
2.	Track the values of each component of the discriminator loss value, 
    generator loss value, and entropy value.
3.	Carefully analyze the performance feedback of the ansatz.
4.	Provide a new and improved ansatz that can better solve the task.

Analysis Criteria:

1.	Discrepancy in Loss Values:
    - If either the discriminator_loss_values or generator_loss_values decreases 
    significantly while the other does not, the learning is not successful, and 
    the ansatz may need to be completely redesigned.
    - If there is a significant discrepancy between the discriminator_loss_values
    and generator_loss_values, the learning is also not successful, indicating 
    a potential need to redesign the ansatz from scratch.
    - Conversely, if both the discriminator_loss_values and generator_loss_values 
    are relatively small and converge to similar values, the ansatz is likely effective.
2.	Convergence of Entropy and KS Test Values:
    - Both the entropy_values and kstest_values should ideally be small. I
    f they do not converge well or exhibit significant oscillations, it indicates that 
    the ansatz configuration is suboptimal, and changing the entanglement strategy might 
    be beneficial.
3.	Parameters and Depth:
    - The number of parameters ({circuit parameter counts}) and the depth ({circuit depth})
    of the ansatz should be minimized for practical implementation on actual devices. 
    While improving the entropy and loss values is important, reducing the parameters 
    and depth is also crucial. For instance, if the entropy and loss values are similar 
    across different ansatzes, the one with fewer parameters and lesser depth should be 
    chosen.

    
Based on the above analysis criteria, provide the code for the improved ansatz.

==INSTRUCTION ==

Please output an ansatz number as well as the selected qubits for each block. 
For example, [1,2,3,4] means we use operation 1 for block 1 and the block 1 is 
on all the qubits, operation 2 for block 2 and block 2 is on all the qubits, 
and so on, until operation 4 for block 4 and block 4 is on all the qubits.
Also, you can choose the same ansatz even if you have already selected the 
ansatz for the previous block.

Please output the values like this:

```
improved_ansatz_list = [4,1,5,1]
```


\end{verbatim}
\end{mdframed}
\end{center}

\label{sec:appendices}

\end{document}